\documentclass[letterpaper, 10 pt, conference]{ieeeconf}  % Comment this line out if you need a4paper
\usepackage{graphicx,graphics}
\usepackage{amsmath,amsfonts,mathrsfs,amssymb,dsfont}
\usepackage{booktabs}
\usepackage{cite}
\usepackage{nomencl}\usepackage[labelfont=bf,labelsep=period,justification=justified]{caption}
\usepackage{bbm}
\usepackage{color}
\usepackage{arydshln}

\usepackage{flushend}

\IEEEoverridecommandlockouts                              % This command is only needed if 
\title{\LARGE \bf
Prediction and Control of Projectile Impact Point using Approximate Statistical Moments}

\author{Cenk Demir$^{1}$, Abhyudai Singh$^{2}$
\thanks{$^{1}$C. Demir is with Department of Electrical and Computer Engineering, University of Delaware, Newark, DE USA 19716. {\tt\small cdemir@udel.edu}}
\thanks{$^{2}$A. Singh is with the Department of Electrical and Computer Engineering, Biomedical Engineering, Mathematical Sciences, Center for Bioinformatics and Computational Biology, University of Delaware, Newark, DE USA 19716.
{\tt\small absingh@udel.edu}}}

\begin{document}

\maketitle
\thispagestyle{empty}
\pagestyle{empty}

%%%%%%%%%%%%%%%%%%%%%%%%%%%%%%%%%%%%%%%%%%%%%%%%%%%%%%%%%%%%%%%%%%%%%%%%%%%%%%%%
\begin{abstract}
In this paper, trajectory prediction and control design for a desired hit point of a projectile is studied. Projectiles are subject to environment noise such as wind effect and measurement noise. In addition, mathematical models of projectiles contain a large number of important states that should be taken into account for having a realistic prediction. Furthermore, dynamics of projectiles contain nonlinear functions such as monomials and sine functions. To address all these issues we formulate a stochastic model for the projectile. We showed that with a set of transformations projectile dynamics only contains nonlinearities of the form of monomials. In the next step we derived approximate moment dynamics of this system using mean-field approximation.  Our method still suffers from size of the system. To address this problem we selected a subset of first- and second-order statistical moments and we showed that they give reliable approximations of the mean and standard deviation of the impact point for a real projectile. Finally we used these selected moments to derive a control law that reduces error to hit a desired point.
\end{abstract}
%%%%%%%%%%%%%%%%%%%%%%%%%%%%%%%%%%%%%%%%%%%%%%%%%%%%%%%%%%%%%%%%%%%%%%%%%%%%%%%%

\section{Introduction}
% no \IEEEPARstart
In a broad sense, a projectile is a ranged weapon that moves through the air in the presence of external forces and endures its motion because of its own inertia. Begining from Aristotle and later Galileo, projectile trajectory prediction and its impact points have been studied \cite{hussey1983physics,galileo}. The goal of these studies is to provide a realistic mathematical model in order to explain the behavior of projectiles. Mathematical modeling allows us to understand the motion of projectiles while requiring a much lower cost than experimental analysis. However, mathematical models become rapidly convoluted when all the parameters of the system are considered (e.g. atmospeheric conditions, air density, fuel, etc.). Hence, forming an optimal model, that contains only  critical information about a projectile, is essential. Various approximations, methods, and assumptions are used to obtain reliable models \cite{mavris1998stochastic, charters1955linearized, guidos2002linearized ,zhao2015finite}. In this paper, we define a novel model to predict impact points for ballistic targets. Predicting these points is essential in order to avoid hitting constrained areas \cite{rogers2015stochastic}, and to create effective defense systems for those points \cite{finn1986value}.

In order to reach to a reasonable Impact Point Prediction (IPP), several deterministic and stochastic approaches are studied, such as various types of Kalman Filters \cite{fieee2005comparison}, Maximum Likelihood Estimator \cite{farina2006classification}, and Stochastic Model Predictive Control \cite{rogers2015stochastic}. Recent works have also included the wind effect which is crucial to obtaining realisctic IPP \cite{yuan2014impact}. However, it is highly inaccessible to receive information about wind instantaneously and hastily because of limited sensor accuracy \cite{yanushevsky2007modern}. Furthermore, the deviations of wind speed and direction are random which makes it unrealizable to predict the future evolution of wind \cite{manwell2010wind}. This noise also shifts the impact points of projectiles which makes classic IPP model erroneous.

In this paper, the effect of wind is modelled as a stochastic process. In the presence of such randomness, statistical moments are reliable tools that give us useful information about the mean and variance of impact points. However, it is not always possible to determine statistical moments for highly nonlinear systems because of the unclosed moment dynamics which  means that higher order moments appear in lower ones. To interpret moments in this case, various approximation techniques are used which are called closure techniques (see, e.g., \cite{soltani2015conditional,lee2009moment,gillespie2009moment,kuehn2016moment,singh2011approximate}). Unfortunately, these methods are mainly developed to deal with higher order moments of monomial form. For instance, how to approximate skewness as a function of mean and variance. However, IPP contains nonlinearities of the trigonometric form due to its nature. Here we show that by using Euler formula we can transform a system to new coordinates. The transformed system can be modelled  as classic monomial form with a change of variables. In the next step, we derive moment dynamics for the transformed system. These dynamics are free of trigonometric functions, yet they are still unclosed. Hence, we apply mean-field approximation to close them \cite{Tsarenko1992}. Mean-field approximation gives reliable results in the limit of weak correlation between states of the system, which is the case here due to the presence of independent noise terms in different states of IPP \cite{ghusinga2017approximate}. 

The ultimate aim of aerospace studies is to control projectile around its desired trajectory which is named as projectile guidance. To do so, many guidance laws are developed \cite{zhang2008analysis, metz1986terminal}. In between them, propotional navigation guidance (PNG) is the most used and well performed when the target is stationary \cite{wang2013stochastic}. PNG is applied to a projectile by changing forces that act on it. Such changes can be obtained by using configurations of canards, wings or tails \cite{yanushevsky2007modern}. Change of configurations is achieved through various control strategies \cite{rogers2015stochastic, gross2014impact, costel97pot}. In this paper, we used feedback control to move along a projectile in a desired trajectory. In the next, we start our analysis by defining projectile dynamic models.

\begin{table}[!h]
\vspace{3mm}
	\caption{\label{tab:parameters}Summary of notation used}
	\centering
	\begin{tabular}{cl} 
		\hline
		Parameter &Description\\
		\hline
$x, y, z$& Location of projectile\\
$\theta, \psi, \phi$& Roll, pitch, yaw angles\\
$p, q, r$ &Angular velocity components\\
$u, v, w$& Projectile velocity components\\
$X, Y, Z$& Force components of projectile\\
$L, M, N$& Physical moment components of projectile\\
$I$& Inertia matrix\\
$s_\theta, c_\theta, t_\theta$& Sine, cosine and tangent of $\theta$\\
$V$& Total velocity\\
$D$&Projectile characteristic length\\
$\rho$ &Atmospheric density\\
$C_{NA}$& Normal force aerodynamic coefficient\\
$v_w, w_w$& Wind components\\
$C_{LP}, C_{MQ}$& Roll and pitch damping aerodynamic coefficients\\
$R_{MCM}$& Distance from center of mass to center of\\
&  Magnus pressure\\
$R_{MCP}$&Distance from center of mass to center of pressure\\
$C_{YPA}$& Magnus force aerodynamic coefficient\\
$C_{x0}$& Zero yaw axial force aerodynamic coefficient\\
$W_1, W_2, W_3$& Zero-mean white noise components\\
$\alpha, \beta$& Angle of attack and side slip angle \\
$u_{ci}, v_{ci}, w_{ci}$& $i$th canard velocity \\
$D_i$& $i$th canard characteristic length \\
$S_i$& $i$th canard surface area \\
$L_i, D_i$& $i$th canard lift and drag force \\
$C_L, C_D$& Canard aerodynamic lift and drag coefficients\\
$M_i$& $i$th canard Mach number \\
$\lambda_i$& $i$th canard angle \\
$r_{xi}, r_{yi}, r_{zi}$& $i$th canard distance from center of gravity \\
$\theta_E, \psi_E$& Yaw and pitch angle error \\
$K_p, K_{\psi}, K_{\theta}, K_{\phi}$& Control gains \\
$e_p, e_q, e_r$& Control inputs \\
		\hline
	\end{tabular}
\end{table}

\section{Exact Dynamic Model of Projectile}

\begin{figure}[!b]
\centering
\includegraphics[width=1\columnwidth]{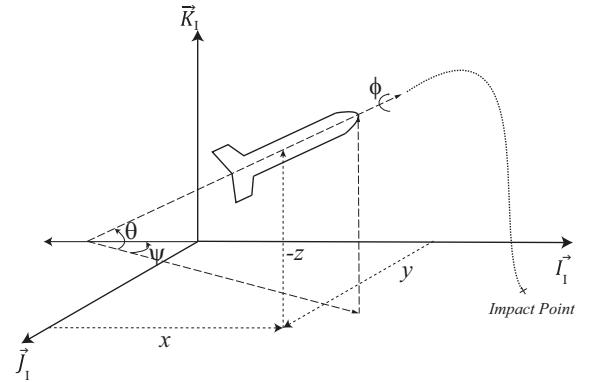}
\caption{\textbf{Model schematic of a projectile in inertial coordinate frame.} A projectile is fully identified by its position in inertial coordinates ($x, y, z$) and its orientation ($\phi, \theta, \psi$). By knowing these 6 degrees of freedom one can predict the impact points, i.e. the point where the projectile hits.}
\label{fig:6DOF}
\end{figure}

To model a projectile motion we first need to define proper coordinates. Projectile dynamics contain two frames. One of them is the inertial frame which is often construed as the Earth coordinate frame. The other one is the projectile referenced frame which is the body frame. In these frames, position states are denoted by ($x, y, z$), and orientation states are denoted by ($\psi, \phi, \theta$). Note that position ($x, y, z$) and orientation ($\phi, \theta, \psi$) are independent parameters, and called six degrees of freedom \cite{mccoy1999modern,amoruso1996euler} which can also be seen in Figure \ref{fig:6DOF}.

The position states ($x, y, z$) give information about location of the projectile in the Earth coordinate which is portrayed as

\begin{equation}
\begin{Bmatrix}
\dot{x} \\ \dot{y} \\ \dot{z} 

\end{Bmatrix}
=
\begin{bmatrix}
c_\theta c_\psi & s_\phi s_\theta c_\psi-s_\psi c_\phi & s_\theta c_\psi c_\phi + s_\psi s_\phi \\
c_\theta s_\psi & s_\phi s_\theta s_\psi+c_\psi c_\phi & s_\theta s_\psi c_\phi - c_\psi s_\phi \\

-s_\theta & s_\phi c_\theta & c_\theta c_\phi
\end{bmatrix}
\begin{Bmatrix}
u\\v\\w
\end{Bmatrix}.
\label{eqn:n1}
\end{equation}
This equation shows that position states ($x, y, z$) depend on the body translational velocities ($u, v, w$). Furthermore $x, y, z$ depend on angles between projectile and Earth frame origin (Figure \ref{fig:6DOF}). The matrix in equation \eqref{eqn:n1} is known as rotation matrix. This matrix is created by the standard aerospace rotation sequence \cite{fresconi2014theory}. It allows us to represent the speed of projectile in all directions in the Earth coordinate system.

The orientation states of the projectile, ($\phi, \theta, \psi$), are known as Euler's angles, they are called specifically: roll, pitch, and yaw. The inertial angular rates ($p, q, r$) are the time derivative of Euler angles ($\dot{\phi}, \dot{\theta},\dot{\psi}$) 
\begin{equation}
\begin{Bmatrix}
p\\q\\r
\end{Bmatrix}=\boldsymbol{L_E^B}
\begin{Bmatrix}
\dot{\phi} \\ \dot{\theta} \\ \dot{\psi}
\end{Bmatrix}.
\label{eqn:ortn}
\end{equation}
This transformation is a non-orthogonal transformation because the axes on which we measured Euler angle derivatives are non-orthogonal. For this reason, it is necessary to define $\boldsymbol{L_E^B}$ which is the matrix that allow us to separate Euler angle derivatives to orthonormal  components \cite{stengel2015flight}. Thus, from taking inverse of $\boldsymbol{L_E^B}$, we can define the dynamics of $\dot{\phi}, \dot{\theta},\dot{\psi}$ 
\begin{equation}
\begin{Bmatrix}
\dot{\phi} \\ \dot{\theta} \\ \dot{\psi}
\end{Bmatrix}
=
\begin{bmatrix}
1& s_{\phi} t_{\theta} & c_{\phi} t_{\theta} \\
0& c_{\phi} & -s_{\phi} \\
0& \frac{s_{\phi}}{c_{\theta}} &\frac{c_{\phi}}{c_{\theta}}
\end{bmatrix}
\begin{Bmatrix}
p\\q\\r
\end{Bmatrix}.
\label{eqn:n2}
\end{equation}

The body velocity components depend on the acting force on the projectile ($X, Y, Z$); explicitly gravitational force, aerodynamic steady state force, aerodynamic Magnus force, and canard lifting force \cite{costello2001extended}. Further, velocity components ($u, v, w$) are 
\begin{equation}
\begin{Bmatrix}
\dot{u}\\ \dot{v} \\ \dot{w}
\end{Bmatrix}
=
\begin{Bmatrix}
\frac{X}{m}\\ \frac{Y}{m} \\ \frac{Z}{m}
\end{Bmatrix}
-
\begin{bmatrix}
0& -r & q \\ r & 0 & -p \\ -q & p & 0
\end{bmatrix}
\begin{Bmatrix}
u \\ v \\ w
\end{Bmatrix}.
\label{eqn:n3}
\end{equation}
Moreover, angular velocity depends on the physical moments that acting on a projectile, ($L, M, N$). These moments are aerodynamic steady state moment, aerodynamic unsteady moment, aerodynamic Magnus moment, and the moment of the canard \cite{costello2001extended}. The connection between angular velocity and physical moment is given by 
\begin{equation}
\begin{Bmatrix}
\dot{p}\\ \dot{q} \\ \dot{r}
\end{Bmatrix}
=
\begin{bmatrix}
I
\end{bmatrix}^{-1}
\begin{bmatrix}

\begin{Bmatrix}
L\\ M \\ N
\end{Bmatrix}
-
\begin{bmatrix}
0& -r & q \\ r & 0 & -p \\ -q & p & 0
\end{bmatrix}
\begin{bmatrix}
I
\end{bmatrix}
\begin{Bmatrix}
p\\q\\r
\end{Bmatrix}
\end{bmatrix}
\label{eqn:n4},
\end{equation}
where $I$ is inertia matrix of a projectile
\begin{equation}
I=
\begin{bmatrix}
I_{XX}& I_{XY} & I_{XZ} \\ I_{YX} & I_{YY} & I_{YZ} \\ I_{ZX} & I_{ZY} & I_{ZZ}
\end{bmatrix}.
\label{eqn:n5}
\end{equation}
The force and the moment terms in equations \eqref{eqn:n3} and \eqref{eqn:n4} are defined as the following: the gravitational force is the force that pulls projectile towards the center of the Earth \cite{cook2012flight}. Aerodynamic force and physical moment are the effect of pressure and shear stress on the body of projectile \cite{anderson2010fundamentals}. This force can be split into lift force which is  opposite to gravitational force, and drag force which is perpendicular to gravitational force and opposite to the lateral velocity of a projectile. After these definitions, the aerodynamic moment can be constructed for any point on the body by using these forces. In addition, canard lifting force and moment are the aerodynamic force and moment over the canard surface. The last phenomena on the projectile is Magnus force and moment which is seen in the spinning projectiles. The pressure difference between opposite sides of a spining projectile creates this force and moment \cite{mccoy1999modern}. 

Equations  \eqref{eqn:n1}-\eqref{eqn:n5} are full description of dynamics for a projectile \cite{etkin2012dynamics}, and it is clear that these dynamics are highly nonlinear. Hence different variations of this model are used by considering different assumptions. In the next section, we briefly review the current approximations.

\section{Linear Model and Modified Linear Model of Projectile}
In order to decrease the complexity of projectile dynamics, researchers came up with different assumptions \cite{etkin2012dynamics, murphy1963free}. One famous framework which is built based on series of simplifying assumptions is known as projectile linear theory. This theory allows us to define analytic solution of projectile \cite{costello2004flight}, however it usually generates considerable error on impact point prediction \cite{hainz2005modified}. To overcome this issue, projectile modified linear model is interpreted with relaxing some of the assumptions. Specifically,
\begin{itemize}
\item The pitch angle, $ \theta $, is not small
\begin{equation}
\sin{(\theta)}\neq\theta, \ \cos{(\theta)}\neq 1.
\end{equation}
\item Projectile roll rate and pitch angle are not assumed constant
\begin{equation}
p\neq p_0, \\ \	 \theta\neq\theta_0.
\end{equation}
\end{itemize} 
After these two changes projectile modified linear model is
\begin{small}
\begin{gather}\small
\dot{x}=\cos{(\theta)} D, \\
\dot{y}=\cos{(\theta)} D \psi + \frac{D}{V} \tilde{v}, \\
\dot{z}=-D \sin{(\theta)} + \frac{D \cos{(\theta)}}{V} \tilde{w},
\\
\dot{\phi}=\frac{D}{V} \tilde{p},\\
\dot{\theta}=\frac{D}{V} \tilde{q}, \\
\dot{\psi}=\frac{D}{V \cos{(\theta)}} \tilde{r}, \\
\dot{V}=-\frac{\pi \rho D^3 C_{x0} V}{8 m} - \frac{D g \sin{(\theta)}}{V}, 
\end{gather} 
\end{small}
\begin{small}
\begin{gather}	
\dot{\tilde{v}}=-\frac{\pi \rho D^3 C_{NA} (\tilde{v} - \tilde{v}_w)}{8 m} - D \tilde{r}, \\
\dot{\tilde{w}}=-\frac{\pi \rho D^3 C_{NA} (\tilde{w} - \tilde{w}_w)}{8 m} + D \tilde{q} + \frac{D g \cos{\theta}}{V}, 
\\
\dot{\tilde{p}}=\frac{\pi \rho D^4 V C_{DD}}{8 I_{xx}} + \frac{\pi \rho D^5 C_{LP} \tilde{p}}{16 I_{xx}}, \\
\dot{\tilde{q}}=\frac{\pi \rho D^4 R_{MCM} C_{YPA} \tilde{p} (\tilde{v} -\tilde{v}_w)}{16 V I_{yy}} \nonumber + \frac{\pi \rho D^5 C_{MQ} \tilde{q}}{16 I_{yy}} \\ + \frac{\pi \rho D^3 R_{MCP} C_{NA} (\tilde{w} - \tilde{w}_w)}{8 I_{yy}} - \frac{I_{xx} D \tilde{p} \tilde{r}}{I_{yy} V}, \\
\dot{\tilde{r}}=\frac{\pi \rho D^4 R_{MCM} C_{YPA} \tilde{p} (\tilde{w} -\tilde{w}_w)}{16 V I_{yy}} \nonumber + \frac{\pi \rho D^5 C_{MQ} \tilde{r}}{16 I_{yy}}\\ - \frac{\pi \rho D^3 R_{MCP} C_{NA} (\tilde{v} - \tilde{v}_w)}{8 I_{yy}} + \frac{I_{xx} D \tilde{p} \tilde{q}}{I_{yy} V}
\end{gather} 
\end{small}
\cite{hainz2005modified}. 
Note that in this paper, a superscript $\sim$ is used to describe  components in the fixed frame.

All of these projectile models are deterministic. However, in reality, motion of projectiles is not deterministic, and has noise components. This calls for a new model for a projectile. In the next section, we will introduce a new stochastic model for a projectile with stochastic noise effect.

\section{Stochastic Modified Linear Model of Projectile}

We start with introducing the general Stochastic Differential Equation (SDE). A typical equation of the form of SDE can be written as   
\begin{equation}
d\boldsymbol{x}=\boldsymbol{f}(\boldsymbol{x},t)dt+\boldsymbol{g}(\boldsymbol{x},t)d\boldsymbol{W}_t,
\label{eqn:SDE}
\end{equation}
where $\boldsymbol{x}=[x_1 \ x_2 \ ... \ x_n]^T \in \mathbb{R}^n$ is the state vector. The system dynamics are $\boldsymbol{f}(\boldsymbol{x},t)=[f_1(\boldsymbol{x},t) \ f_2(\boldsymbol{x},t) \  ... \ f_n(\boldsymbol{x},t)]^T \in \mathbb{R}^n \times [0,\infty ) \rightarrow \mathbb{R}^n$, and $\boldsymbol{g}(\boldsymbol{x},t)=[g_1(\boldsymbol{x},t) \ g_2(\boldsymbol{x},t) \  ... \ g_n(\boldsymbol{x},t)]^T \in \mathbb{R}^n \times [0,\infty ) \rightarrow \mathbb{R}^n$. $\boldsymbol{W}_t=[W_1 W_2 ... W_n]$ is a $n$-dimensional Wiener process whose mean is zero. 

Projectiles contain miscellaneous physical components that are subject to noise. Various noise sources include measurement noise, sensor noise, wind effect and, etc. Consistent with previous studies, we model these noise terms as state independent white noise \cite{yuan2014impact, yuan2010impact, hutchins1998imm, tsai1991angle, maley2015optimal}
\begin{gather}
dx=(\cos{(\theta)} D)dt + a_1 d{W_1}, \label{eqn:noise1} \\ 
dy=(\cos{(\theta)} D \psi + \frac{D}{V} \tilde{v})dt + a_2 d{W_2}, \label{eqn:noise2} \\ 
dz=(-D \sin{(\theta)} + \frac{D \cos{(\theta)}}{V} \tilde{w})dt + a_3 d{W_3}, \label{eqn:noise3}
\end{gather}
where $a_1$, $a_2$ and $a_3$ are constant, and 
\begin{gather}
\langle dW_i dW_i \rangle = dt, \ \ \ \
\langle dW_i dW_j \rangle = 0, i \neq j
\end{gather}
In the rest of this paper, $\langle \rangle$ denotes the expected value. $W_1, W_2$ and $W_3$ can be understood as the effect of noise on different coordinates (Figure \ref{fig:9tra}). After these modifications, dynamics of projectile is in the form of SDE equation \eqref{eqn:SDE}, thus we call this model Projectile Stochastic Modified Linear Model (PSMLM).

\subsection{Statistical Moment Dynamics of PSMLM}
The moment dynamics of a general SDE in the form of equation \eqref{eqn:SDE} can be derived by using It\o \ formula
\begin{small}
\begin{equation}
\frac{d\langle \boldsymbol{x^{[m]}} \rangle}{dt}= \sum_{i=1}^n \bigg \langle f_i \frac{\partial \boldsymbol{x}^{[m]}}{\partial x_i} \bigg \rangle + \frac{1}{2} \sum_{i=1}^n \sum_{j=1}^n \left( (\boldsymbol{g} \boldsymbol{g}^T)_{ij}\frac{\partial^2 \boldsymbol{x}^{[m]}}{\partial x_i \partial x_j} \right) \bigg\rangle,
\label{eqn:Ito}
\end{equation} 
\end{small}
where $[m]=[m_1 \ m_2 \ ... m_n]^T$ \cite{hespanha2005stochastic}.
The sum of $m_j$ is the order of the moment. If $\boldsymbol{f}(\boldsymbol{x},t)$ and $\boldsymbol{g}(\boldsymbol{x},t)$ are linear, then moment dynamics can be written compactly as
\begin{equation}
\frac{d \boldsymbol{\mu}}{dt}=\boldsymbol{c}+\boldsymbol{A \mu},
\label{eqn:master}
\end{equation}
where $\boldsymbol{\mu}$ contains all the moment of the system up to order M 
\begin{gather}
\text{M}\equiv \sum_{j=1}^{n} m_j.
\end{gather}
The vector $\boldsymbol{c}$, and the matrix $\boldsymbol{A}$ are determined by using $\boldsymbol{f}(\boldsymbol{x},t)$ and $\boldsymbol{g}(\boldsymbol{x},t)$. To solve the linear equation \eqref{eqn:master} is effortless because the desired moment order is always combination of higher or the same order moments. However, when $\boldsymbol{f}(\boldsymbol{x},t)$ or $\boldsymbol{g}(\boldsymbol{x},t)$ are nonlinear, it   is not simple to determine all the moments, and \eqref{eqn:master} needs new configuration. This is 
\begin{equation}
\frac{d \boldsymbol{\mu}}{dt}=\boldsymbol{c}+\boldsymbol{A \mu}+\boldsymbol{B \bar{\mu}},
\label{eqn:master2}
\end{equation}
where $\boldsymbol{\bar{\mu}}$ only contains the higher moments than the desired ones.

This problem is a fundamental problem of statistical moments determination when $\boldsymbol{f}(\boldsymbol{x},t)$ or $\boldsymbol{g}(\boldsymbol{x},t)$ are nonlinear. To overcome this fundamental problem, we used well-known mean-field closure technique  \cite{bobbio2008mean, chibbaro2014stochastic}. This approach is convenient for the systems that have computational complexity or high-dimension \cite{vrettas2015variational}. The basic idea of this moment closure technique is to define higher order moments as the product of the moments of individuals. For instance
\begin{equation}
\langle x_i^{m_i} x_j^{m_j} \rangle\approx\langle x_i^{m_i} \rangle \langle x_j^{m_j} \rangle,\ \ \ \  m_i, m_j \in [0, 1, 2, \ldots].
\label{eqn:independency}
\end{equation}

\subsubsection{Approximation of mean of PSMLM using mean-field}
It is handy to use \eqref{eqn:Ito} to find moments of the system. For example for state $x$,
\begin{equation}
dx=\frac{D}{2}\cos{(\theta)}dt + a_1 d{W_1}.
\label{eqn:mean1}
\end{equation}
This dynamic is nonlinear because of $cosine$ term. If we use Euler's relation, 
\begin{gather}
\cos{\theta}=\frac{e^{i \theta}+e^{-i \theta}}{2}, 
\ \
\sin{\theta}=\frac{e^{i \theta}-e^{-i \theta}}{2 i}.
\end{gather}
Thus, \eqref{eqn:mean1} can be rewritten as
\begin{equation}
dx=(\frac{D}{2}(e^{j \theta}+e^{-j \theta})dt + d{W_1}.
\label{eqn:xdyn}
\end{equation}
To proceed from \eqref{eqn:xdyn}, we define two new states $\delta^+=e^{j \theta}$ and $\delta^-=e^{-j \theta}$
\begin{gather}
d\delta^+=j \frac{D}{V_0} \delta^+,  \ \ \ \ 
d\delta^-=-j \frac{D}{V_0} \delta^-
\label{eqn:e2}
\end{gather}
We start our analysis by writing the first statistical moment dynamics,
\begin{gather}\small
	\frac{d\langle \delta^+ \rangle }{dt}=j 				\frac{D}{V_0} \langle \delta^+ \tilde{q} 				\rangle  \label{eqn:ei}\\ 
	\frac{d\langle e^{-j \theta}\rangle }{dt}=-j			\frac{D}{V_0} \langle \delta^- \tilde{q} 				\rangle \label{eqn:-ei}\\
	\frac{d \langle x \rangle}{dt}=(\frac{D}{2}( 			\langle \delta^+ \rangle + \langle \delta^- \rangle ) \label{eqn:xcor}
\end{gather}
These moment dynamics depend on the second order moment dynamics $\langle \delta^+ \tilde{q} \rangle$ and $\langle \delta^- \tilde{q} \rangle$. In the next step, we add dynamic of these two moment to our system which are
\begin{small}
\begin{gather}
	\frac{d \langle \delta^+ \tilde{q} \rangle}{dt}=  j \frac{D}{V_0} \langle \delta^+ \tilde{q}^2 	\rangle  + \frac{\pi \rho D^4 R_{MCM} 				C_{YPA}}{16 V_0 I_{yy}} \langle \delta^+ \tilde{p} (\tilde{v} -\tilde{v}_w) \rangle \nonumber \\  + \frac{\pi \rho D^5 C_{MQ}}{16 I_{yy}} \langle \delta^+ \tilde{q} \rangle + \frac{\pi \rho D^3 R_{MCP} C_{NA}} {8 I_{yy}} \langle \delta^+ (\tilde{w} - \tilde{w}_w) \rangle \nonumber \\  - \frac{I_{xx} D }{I_{yy} V_0} \langle \delta^+ \tilde{p} \tilde{r}  \rangle, \label{eqn:eij} \\
		\frac{d \langle \delta^- \tilde{q} \rangle}{dt}=  -j \frac{D}{V_0} \langle \delta^- \tilde{q}^2 \rangle  + \frac{\pi \rho D^5 C_{MQ}}{16 I_{yy}} \langle \delta^- \tilde{q} \rangle \nonumber \\ \nonumber +  \frac{\pi \rho D^4 R_{MCM} C_{YPA}}{16 V_0 I_{yy}} \langle \delta^-  \tilde{p} (\tilde{v}-\tilde{v}_w) \rangle \nonumber \\ + \frac{\pi\rho D^3 R_{MCP} C_{NA}}{8 I_{yy}} \langle \delta^- (\tilde{w} - \tilde{w}_w) \rangle - \frac{I_{xx} D}{I_{yy} V_0} \langle \delta^- \tilde{p} \tilde{r} \rangle. \label{eqn:-eij}
\end{gather}
\end{small}
These dynamics depend on the third order moments $\langle \delta^+ \tilde{q}^2 \rangle$ and $\langle \delta^- \tilde{q}^2 \rangle$. The moment dynamics of this system are not closed in the sense that any close set of moments depends on higher order moments. To overcome this issue, mean-field closure technique is used 
\begin{gather}
\langle \delta^+ \tilde{q}^2 \rangle \approx \langle \delta^+ \rangle \langle \tilde{q}^2 \rangle, ~ \langle \delta^- \tilde{q}^2 \rangle \approx \langle \delta^- \rangle \langle \tilde{q}^2 \rangle.
\end{gather}
In order to calculate mean, we need to add second order moment dynamics in the next.

\begin{figure}[!b]
\centering
\includegraphics[width=1 \columnwidth]{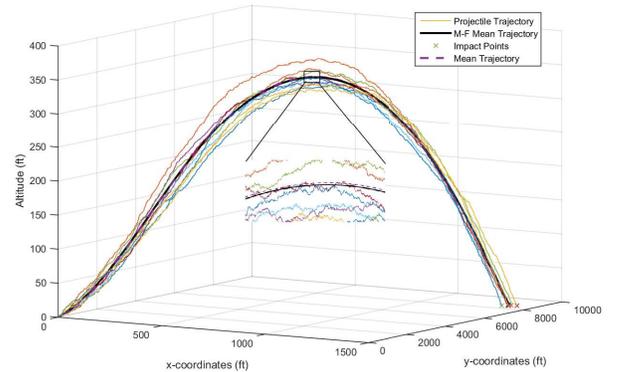}
\caption{\textbf{Our approximation technique captures the mean behavior of projectile}. The bold lines show the mean trajectory of a projectile obtained from numerical simulations and mean-field approximation. The approximation results are indistinguishable from numerical results, showing significant performance of our method.}
\label{fig:9tra}
\end{figure}
\subsubsection{Second order moments of PSMLM}
This system has 14 states, which means there exist 14 first order statistical moments, and 105 second order statistical moments. This number of equations make running different methods and finding optimal solutions in real time impossible. Hence, here we only consider a few number of second order moments and we approximate the rest as functions of first order moments. 

Namely we add the 14 equations of the form of a sole state, for instance
\begin{equation}
\frac{d \langle x^2 \rangle}{dt}= D ( \langle x \delta^+ \rangle + \langle x \delta^- \rangle ) + a_1.
\label{eqn:sec}
\end{equation}
Moreover, we include $\langle x \delta^+ \rangle$, $\langle x \delta^- \rangle$ and the moments that show up in the moment dynamics of coordinates. The rest of moments are approximated, 
\begin{gather}
\langle \delta^+ \tilde{q} 	\rangle \approx \langle \delta^+ \rangle \langle \tilde{q} 	\rangle,  ~ 
	\langle \delta^- \tilde{q} \rangle \approx \langle \delta^- \rangle \langle \tilde{q} \rangle, \\	 
	 \langle \delta^+ \tilde{p} \tilde{v} \rangle \approx \langle \delta^+ \rangle \langle \tilde{p} \tilde{v} \rangle, ~
\langle \delta^-  \tilde{p} \tilde{v} \rangle \approx \langle \delta^- \rangle \langle \tilde{p} \tilde{v} \rangle, \\ 	
\langle \delta^+ \tilde{p} \tilde{r}  \rangle \approx \langle \delta^+ \rangle \langle \tilde{p}  \tilde{r} \rangle,~ \langle \delta^- \tilde{p} \tilde{r} \rangle \approx \langle \delta^- \rangle \langle \tilde{p}  \tilde{r} \rangle.
\end{gather} 
By applying these closures, we have closed set of first order moment dynamics and selected sub-set of second order moments of PSMLM. Solving these equations give us approximate time trend of the mean and the standard deviation.

\begin{figure}[!t]
\centering
\includegraphics[width=1\columnwidth]{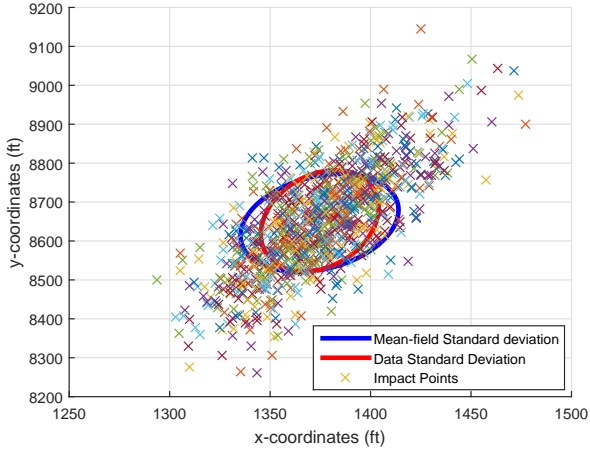}
\caption{\textbf{Our method successfully predicts the standard deviation of the impact points.}
Every cross represents impact point when a projectile have noise component. The blue and red ellipse are  the standard deviation of x and y coordinates by using mean-field approximation and data simulations, respectively.}
\label{fig:1000ipp}
\end{figure}

\begin{figure}[!t]
\centering
\includegraphics[width=1\columnwidth]{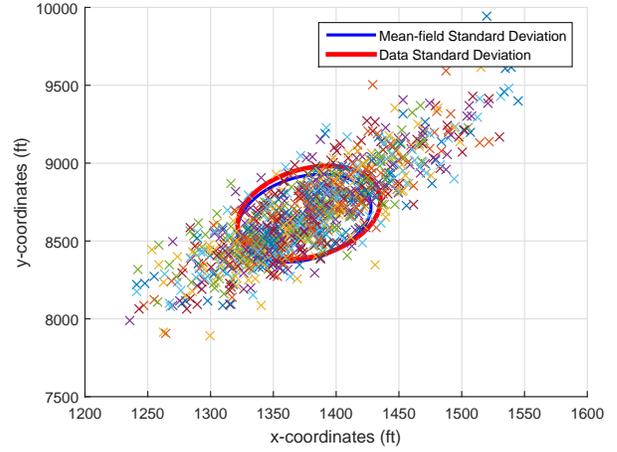}
\caption{\textbf{Randomness in starting point increase the errors in impact points, hence it vitiates the accuracy of the missile.} Mean field approximation is capable of giving accurate estimations of error even in the presence of random initial conditions.}
\label{fig:comparison}
\end{figure}

\section{Simulation Results of Projectile Prediction Using Mean-field}

In this paper, fin stabilized projectile is used with following initial conditions for location, $x=0$ ft, $y=0$ ft, and $z=0$ ft which is origin of reference frame. Speed terms are considered as $\tilde{u}=400$ ft/s, $\tilde{v}=0$ ft/s, and $\tilde{w}=0$ ft/s. Angles in reference coordinates system are $\phi=2.9$ rad, $\theta=0.267$ rad, and $\psi=-0.007$ rad. Angular velocities are $\tilde{p}=399.7$ rad/sec, $\tilde{q}=0.43$ rad/sec, and $\tilde{r}=-1.54$ rad/sec. Physical parameters are selected as air density of $\rho=0.00238$ slug/(cu ft), and gravity constant of $g=32.174$ ft/s$^2$. The rest of physical parameters are chosen as in \cite{costello2004flight}, i.e. reference diameter is $0.343521$ ft, weight of the projectile is $m=0.0116$ slug. Aerodynamic coefficients are $C_{X0}=0.279$, $C_{dd}=2.672$, $C_{lp}=-0.042$, $C_{na}=2.329$, $C_{ypa}=-0.295$, $C_{mq}=-1.800$. Distances from center of mass for aerodynamic moment and force are $R_{mcp}=-0.1657$ ft, and $R_{mcm}=-0.1677$ ft. Moments of inertia are $I_{xx}=2.85\times10^{-5}$ slug/ft$^2$, $I_{yy}=2.72\times10^{-5}$ slug/ft$^2$\cite{costello2004flight}. In addition, wind speed can be taken from weather cast agencies, here we used $\tilde{v}_w=15$ ft/s and $\tilde{w}_w=15$ ft/s.

After using these initial conditions and parameters, we simulated PSMLM. The result of this simulation can be seen in Figure \ref{fig:9tra}. As it is clear from this figure, our approximation is able to predict the mean behavior of projectile through the time successfully. In the next, we analyzed the impact points. Figure \ref{fig:1000ipp} shows that our method is able to predict standard deviation of impact points around their mean with small error. 
\begin{table}[!h]
	\caption{\label{tab:initial}Initial Values of Probability Distribution}
	\centering
	\begin{tabular}{cccccc} 
		\hline
		Parameter  & Mean &S.D. &Parameter &S.D. & Mean \\
		\hline
		$x$& 0&3 & $u$ & 400 &2\\
		$y$& 0&3 &$v$ & 5 & 0.01\\
		$z$& 0&0 &$w$ & 5 & 0.001\\
		$\phi$& 2.9&1 &$p$ & 399.7 &3\\
		$\theta$& 0.267 &0.017 &$q$ &0.43 & 0.01\\
		$\psi$& -0.007 &0.002 &$r$ & -1.54 & 0.01\\
 \hline
	\end{tabular}
\end{table} 

Moreover, in reality the initial condition may change because of the randomness in nature such as wind, and topology of the terrain. To address such uncertainty projectile prediction is performed for a distribution of initial conditions \cite{rogers2015stochastic, ollerenshaw2008model, rogers2013design}. We implemented this source of uncertainty in our numerical simulations by choosing a random initial condition drawn from a distribution introduced in Table \ref{tab:initial}. Figure \ref{fig:comparison} shows that our method is still capable of predicting impact points. As expected standard deviation increases in presence of stochastic starting point.

\section{Design of Projectile Control Law}

In this section, we focus on controlling a projectile. To apply control, four canards are used whose characteristics are exactly the same. These canards allow us to adjust forces and physical moments that act on a projectile \cite{costel97pot}. Controlling these forces and moments are generally obtained by manipulating canard angles ($\lambda_1, \lambda_2, \lambda_3, \lambda_4$). Here, we design a feedback control with the subjective of following a desired trajectory (e.g. the trajectory with minumum flight time \cite{yanushevsky2007modern}) by using these angles. In the next, we define canard properties and feedback control law with details. 

The four symmetric canards are added into projectile.  The lift force of canard is \cite{mccoy1999modern}
\begin{gather}
L_i=\frac{1}{2} \rho (u_{ci}^2+v_{ci}^2+w_{ci}^2)\frac{\pi D_i^2}{4} C_L, \ i={1,2,3,4},
\label{eqn:can1}
\end{gather}
and the drag force of canard is \cite{mccoy1999modern}
\begin{gather}
D_i=\frac{1}{2} \rho (u_{ci}^2+v_{ci}^2+w_{ci}^2)\frac{\pi D_i^2}{4} C_D, \ i={1,2,3,4},
\end{gather}
where $S_i=\frac{\pi D_i^2}{4}$, and coefficients are defined as \cite{costel97pot} 
\begin{gather}
C_L=c_{l \alpha}(M_i)\alpha_i, \\
C_D=c_{d 0}(M_i)+c_{d2} (M_i) \alpha_i^2 + c_i(M_i) C_L^2.
\end{gather}
Here $M_i$ is Mach number on which aerodynamic coefficients depend \cite{costel97pot}. Mach number is the ratio between speed of sound and the speed of air vehicles \cite{stengel2015flight}. In addition, $\alpha_i$ is the angle of attack of each canard which depends on the location of that canard and projectile angle of attack. For each canard, this angle can be defined as
\begin{gather}
\alpha_i=\lambda_i \pm \tan^{-1}(\frac{w_{ci}}{u_{ci}}).
\end{gather}

We have $4$ canards in different locations. Two of them are located parallel to missile  body in y-direction. The force terms for these two are
\begin{equation}\small
\begin{Bmatrix}
X_{ci} \\ Y_{ci} \\ Z_{ci} 
\end{Bmatrix}
=
\begin{Bmatrix}
L_i \frac{w_{ci}}{\sqrt{u_{ci}^2+w_{ci}^2}}-D_i \frac{u_{ci}}{\sqrt{u_{ci}^2+w_{ci}^2}} \\ 0 \\ -L_i \frac{u_{ci}}{\sqrt{u_{ci}^2+w_{ci}^2}}-D_i \frac{w_{ci}}{\sqrt{u_{ci}^2+w_{ci}^2}}
\end{Bmatrix},
\ i={1,3}.
\end{equation}
The other two are located parallel to missile body in z-direction. Their force terms are
\begin{equation}\small
\begin{Bmatrix}
X_{ci} \\ Y_{ci} \\ Z_{ci} 
\end{Bmatrix}
=
\begin{Bmatrix}
-L_i \frac{v_{ci}}{\sqrt{u_{ci}^2+v_{ci}^2}}-D_i \frac{u_{ci}}{\sqrt{u_{ci}^2+v_{ci}^2}} \\ L_i \frac{u_{ci}}{\sqrt{u_{ci}^2+v_{ci}^2}}-D_i \frac{v_{ci}}{\sqrt{u_{ci}^2+v_{ci}^2}} \\ 0
\end{Bmatrix},
\ i={2,4}.
\end{equation}
Moreover, physical moments of these 4 canards are defined as
\begin{equation}
\begin{Bmatrix}
L_{ci} \\ M_{ci} \\ N_{ci} 
\end{Bmatrix}
=
\begin{bmatrix}
0 & -r{zi} & r_{yi}\\ r_{zi} & 0 & -r_{xi}\\ -r_{yi} &r_{xi} & 0
\end{bmatrix}
\begin{Bmatrix}
X_{ci} \\  Y_{ci} \\ Z_{ci}
\end{Bmatrix},
\label{eqn:canend}
\end{equation}
where $r_{xi}, r_{yi}$ and $r_{zi}$ are the distance from center of gravity for each canard in each direction. Then, these forces and moments are added into the system dynamic model directly.

\begin{figure}[!t]
\centering
\includegraphics[width=1\columnwidth]{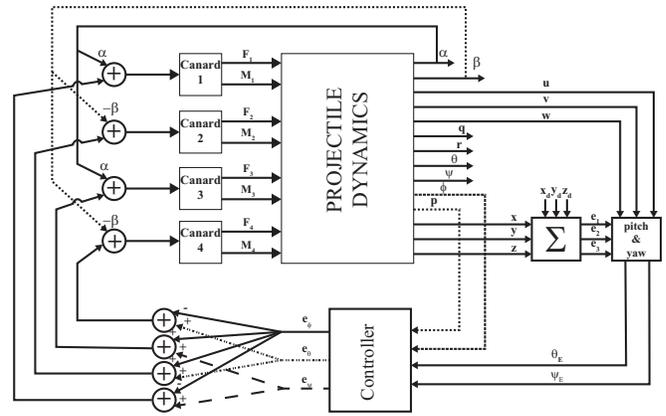}
\caption{\textbf{The controller schematic for following the desired projectile by changing the angle of canards.} The control process starts with getting information about the current states of a projectile ($x, y, z, \alpha, \beta, u, v, w, \phi, p$). The states ($x, y, z$) are subtracted from desired states ($x_d, y_d, z_d$) to find the location error. According to this error, yaw and pitch angle errors are calculated by using velocity states ($u, v, w$). Then, yaw and pitch angle errors and states ($p, \phi$) are used to control system through a feedback. Canards' angle are updated using output of controller, angle of attack and side slip angle.}
\label{fig:cont}
\end{figure}

To design a controller for a projectile by using feedback control law, a desired trajectory is required. In this paper, we assume that the desired trajectory is pre-defined. The error terms ($e_1, e_2, e_3$) between the desired trajectory and actual trajectory are the difference along $x, y$ and $z$ coordinates \cite{costel97pot}. According to these error terms, yaw and pitch angle errors are described as 
\begin{gather}
\theta_{E}=\tan^{-1}(\frac{\tilde{w}}{\tilde{u}})-\tan^{-1}(\frac{e_3}{e_1}), \\
\psi_{E}=-\tan^{-1}(\frac{\tilde{v}}{\tilde{u}})+\tan^{-1}(\frac{e_2}{e_1})
\end{gather}
where $\tan^{-1}(\frac{w}{u})$ and $\tan^{-1}(\frac{v}{u})$ are angle of attack ($\alpha$) and side slip angle ($\beta$).
By using these errors, the control law is 
\begin{gather}
e_{\phi}=K_p p + K_{\phi} \phi, \\
e_{\theta}=K_{\theta} \theta_E, \ \ \ \
e_{\psi}=K_{\psi} \psi_E,
\end{gather}
where $K_p, K_{\phi}, K_{\theta}, K_{\psi}$ are feedback control gains \cite{costel97pot}. Finally, the control is applied by updating the angle of each canard \cite{costel97pot}.
\begin{gather}
\lambda_1=e_{\theta}-e_{\phi}, \ \ \ \
\lambda_2=e_{\psi}+e_{\phi}, \\
\lambda_3=e_{\theta}+e_{\phi}, \ \ \ \
\lambda_4=e_{\psi}-e_{\phi}.
\end{gather}
The entire control design schematic is shown in Figure \ref{fig:cont}.

\section{Simulation Results for Implementing the Controller}
We assumed that the aerodynamic coefficients of canards ($c_{l \alpha}, c_{d 0}, c_{d2}, c_i$) are constant. Also, the distance of each canards from center of gravity is described as the following.

\begin{figure}[!h]
\centering
\includegraphics[width=1\columnwidth]{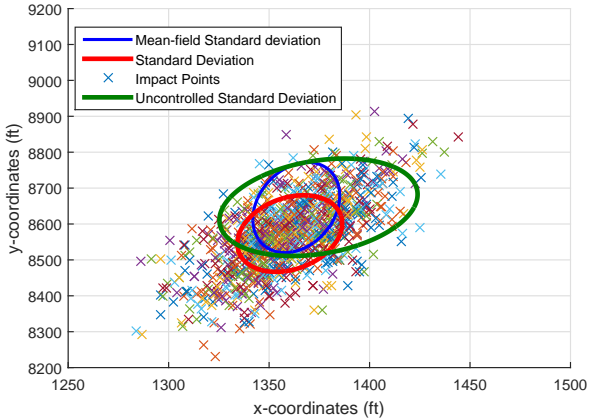}
\caption{\textbf{By controlling a projectile, standard deviation of impact points location reduces considerably.} The control law successfully rejected the contribution of the noise and made the projectile to follow the desired path. This results in lower deviation of impact point.}
\label{fig:control}
\end{figure}

\begin{table}[!h]
	\caption{\label{tab:canarddistance}Distance from center of gravity for each canards}
	\centering
	\begin{tabular}{cccccc} 
		\hline
		Parameter &Value ($ft$) &Parameter &Value &Parameter &Value\\
		\hline
$r_{x1}$& 0.474 & $r_{y1}$& 0.102 & $r_{z1}$& 0\\
$r_{x2}$& 0.474 & $r_{y2}$& 0 & $r_{z2}$& 0.102\\
$r_{x3}$& 0.474 & $r_{y3}$& -0.102 & $r_{z3}$& 0\\
$r_{x4}$& 0.474 & $r_{y4}$& 0 & $r_{z4}$& -0.102 \\
			\hline
	\end{tabular}
\end{table} 	
Because of usage of identical canards, all the canard wing areas are equal to $S_i=0.02104 \ ft^2$. The velocities of canards are not the same because of angular rates and position of each canard \cite{costel97pot}. However, these velocities can be calculated by considering location of canards and angular velocities in each iteration. In control process, we selected the control gain parameters as $K_p=-2, K_{\phi}=-1.5, K_{\theta}=0.01, K_{\psi}=0.015$. By using these control parameters, change of standard deviation and impact points are shown in Figure \ref{fig:control}. Our control law successfully reduced the variance of impact points. Hence, it increases the reliability and accuracy of the missile. However, the error between the simulation results and mean-field approximation is small, but not negligible. One way to reduce this error is to add more moment dynamics to our analysis. Future work will quantify bounds on error of estimation to find the optimal number of dynamics needed to be added to reach to a desired error. Such bounds on approximation error of moments is recently developed for simple dynamic systems \cite{ghusinga2017exact, lamperski2016stochastic}.

\section{Conclusion}
In this paper, we used SDEs to model projectiles under noise effect. Next, we applied Euler's formula to deduce nonlinearities of  the trigonometric to monomial form. Then, we employ mean-field approximation to obtain closed form equations describing mean and standard deviation of the system. Our approximation gives reliable results in predicting time evolution of projectile and characteristics of impact points. Finally, we proposed a control scheme to reduce the errors in impact points.

Furthermore, while the aim of a projectile is to hit the exact target point, it also evades to hit constrained areas. For this purpose, skewness, and kurtosis can be used to avoid hitting those areas by changing the shape of distribution of impact points. Further research will study the higher order moments of projectile such as skewness and kurtosis. Finally in this work we assumed that the controller is built in the projectile. Sometimes we need to give a new control law to the projectile through a transmission channel. Prospect research will merge dynamics of the projectile with random discrete transmission events modelled as renewal transitions to address this requirement \cite{sos17, sos17b, sos16c}.

\section*{Acknowledgment}
\thanks{AS is supported by the National Science Foundation Grant ECCS-1711548.}

%\vspace{-4mm}
%\bibliographystyle{IEEEconf}
\bibliographystyle{IEEEtran}
\bibliography{mybib}
%\bibliographystyle{plain}
%\bibliographystyle{IEEEtran}

%%%%%%%%%%%%%%%%%%%%%%%%%%%%%%%%%%%%%%%%%%%%%%%%%%%%%%%%%%%%%%%%%%%%%%%%%%%%%%%%

\end{document}